\documentclass[rnote]{aa} 
\usepackage{graphicx}
\usepackage[varg]{txfonts}
\usepackage{longtable}

\usepackage{natbib}

\def\ltsima{$\; \buildrel < \over \sim \;$}
\def\simlt{\lower.5ex\hbox{\ltsima}}
\def\gtsima{$\; \buildrel > \over \sim \;$}
\def\simgt{\lower.5ex\hbox{\gtsima}}
\def\AA{\; \buildrel \circ \over {\rm A}}

\begin{document} 
 
\title{A near-ultraviolet view of the Inner Region of M31 with the Large Binocular Telescope}

\author{
G. Beccari\inst{1} \and
M. Bellazzini\inst{1}\and
G. Clementini\inst{1}\and
L. Federici\inst{1} \and
F. Fusi Pecci\inst{1}\and
S. Galleti\inst{1}\and
P. Montegriffo\inst{1}\and
E. Giallongo\inst{2}\and
R. Ragazzoni\inst{3}\and
A. Grazian\inst{2}\and
A. Baruffolo\inst{3}\and
C. De Santis\inst{2}\and
E. Diolaiti\inst{1}\and
A. Di Paola\inst{2}\and
J. Farinato\inst{3}\and
A. Fontana\inst{2}\and
S. Gallozzi\inst{2}\and
F. Gasparo\inst{4}\and
G. Gentile\inst{3}\and
R.Green\inst{5}\and
J. Hill\inst{5}\and
O. Kuhn\inst{5}\and
N. Menci\inst{2}\and
F. Pasian\inst{4}\and
F. Pedichini\inst{2}\and
R. Smareglia\inst{4}\and
R. Speziali\inst{2}\and
V. Testa\inst{2}\and
D. Thompson\inst{5}\and
E. Vernet\inst{6}\and
R.M. Wagner\inst{5}
}
\institute{
INAF, Osservatorio Astronomico di
Bologna, via Ranzani 1, I-40127 
Bologna, Italy;
(giacomo.beccari$@$oabo.inaf.it)
\and
INAF, Osservatorio Astronomico di Roma, Via Frascati 33, I-00040, Monteporzio, Italy
\and
INAF, Osservatorio Astronomico di Padova, Vicolo dell'Osservatorio, 5, I-35122 Padova, Italy
\and
INAF, Osservatorio Astronomico di Trieste, Via G.B. Tiepolo 11, I-34131 Trieste, Italy
\and
Large Binocular Telescope Observatory, University of Arizona, 933 N. Cherry Ave., Tucson, AZ 85721-0065
\and
INAF, Osservatorio Astronomico di Arcetri, Largo E. Fermi 5, I-50125, Firenze, Italy}

\medskip

\abstract
{}
{We present a 900 sec, wide-field $U$ image of the
inner region of the Andromeda galaxy obtained during the commissioning 
of the blue channel of the Large Binocular Camera mounted 
on the prime focus of the Large Binocular Telescope.}
{Relative photometry and absolute astrometry of individual sources in the image
was obtained along with morphological parameters aimed at discriminating between stars and extended sources, e.g. globular
clusters. }
{The image unveils the near-ultraviolet view of the inner ring of star formation recently discovered in the infrared by 
the Spitzer Space Telescope and shows in great detail the fine structure of the dust lanes associated with the
galaxy inner spiral arms. 
The capabilities of the blue channel of the Large Binocular Camera 
at the Large Binocular Telescope (LBC-Blue) are probed by direct comparison
with ultraviolet GALEX observations of the same region in M31. 
We discovered 6 new candidate stellar clusters in this 
high-background region of M31.  We also recovered 62 \textit{bona-fide}  globulars and 62 previously known candidates  
from the Revised Bologna Catalogue of the M31 globular clusters, and firmly established the extended
nature of 19 of them.}
{}

\maketitle

\section{Introduction}
As the nearest giant spiral galaxy, Andromeda (M31) provides a unique opportunity to study the structure 
and evolution of a massive twin galaxy of the Milky Way (MW) from the ``outside'', 
and, by comparison with the MW, to address the question 
of variety in the evolutionary histories of massive spirals. In particular, the M31 Globular Cluster (GC) 
system provides the crucial bridge to connect the integrated-light-based methods that are used to
study GCs in distant galaxies to the resolved-star-based methods that can be applied to the
MW clusters (see e.g. Galleti et al. 2004, 2006a, hereafter G04, G06a).

Hierarchical structure formation models 
predict that the halo of a large galaxy is assembled through mergers of smaller subsystems. 
It has been suggested that M31 originated as an early merger of two or more relatively massive metal-rich 
progenitors (van den Bergh 2000, 2006). 
The prominent tidal stream in M31 (Ibata et al. 2001) extending several degrees from the centre of the 
galaxy (McConnachie et al. 2003), and the newly discovered arc-like over-density connecting M31 to its 
dwarf elliptical companion NGC 205 (McConnachie et al. 2004), are indeed the most spectacular interaction 
signatures presently known in the Local Group (LG) among the few nearby examples of tidal debris trails 
from galaxies currently merging.

The M31 outer disk is warped, and the halo contains numerous loops and ripples.
Moreover, infrared photometry of the disk by IRAS (Habing et al. 1984) and by 
the Spitzer Space Telescope has revealed 
two rings of star formation off-centred from the galaxy nucleus
Gordon et al. (2006).
The two rings appear to be density waves propagating in the disk.
Numerical simulations show that both rings may result from a companion galaxy plunging
head-on through the centre of the M31 disk about 210 Myr ago (Block et al. 2006).
The Galaxy Evolution Explorer (GALEX, Thilker et al. 2005) obtained far 
(1350-1750 $\AA$, FUV) and near (1750-2750 $\AA$, NUV) ultraviolet mosaic imaging 
of the M31 nucleus 
at a spatial resolution of 5$^{\prime\prime}$ in NUV. The M31 inner ring region is not well detected 
by GALEX. 
The Large Binocular Telescope (LBT), a
forefront observational facility built and operated by an
Italian-German-American collaboration, is
at present the only telescope capable to unveil the near UV 
counterpart of the Spitzer inner ring in M31, thanks to the optimal combination of telescope size, spatial 
resolution (a factor 5 higher than in GALEX), field of view and sensitivity in the blue bands.
The LBT is located on Mount Graham, Arizona,
and in its final configuration will have two 8.4 m primary mirrors on a common
mount, feeding various instruments (Salinari 1999,  Hill et al. 2006). 
Science commissioning at prime focus was carried out during October - December 
2006 with a single mirror feeding
the blue channel of the Large
Binocular Camera (LBC-Blue; Ragazzoni et al. 2006, Giallongo et al. 2007)
, a UV-optimized wide-field mosaic camera consisting of four CCD chips of 2048 $\times$ 4608
pixels, with a mean pixel scale of $0.225\arcsec/$pixel. The mosaic covers a field of view (FOV)
of  24$^{\prime} \times 25^{\prime}$; once the inter-chip gaps are taken into account, the effective
sampled area is $\sim 23^{\prime} \times 23^{\prime}$.

In this paper we present results from a 900 sec wide-field $U$ 
band image of the
M31 inner regions, reaching the limiting magnitude $U \sim$ 25 mag. This image was 
obtained with the LBT during the commissioning 
phase, with the purpose of testing the $U$ capabilities of the blue channel of the LBC.
The image is compared to the GALEX and Spitzer views of the same area.
Results are also presented on new M31 candidate stellar clusters that were identified 
on the image.

\section{Observations and data analysis}

The $t_{exp}=900$ s $U$ (Bessel) image of M31 that is the subject of the present analysis was
acquired on UT 2006,  December 25, with the LBT telescope set at RA(J2000)=00 42 42.8 and 
DEC(J2000)=+41 16 12.8.
The selected field is centred on the
M31 inner dust ring discovered by Spitzer at about 0.5 kpc from the M31 centre.
The inner ring extends for about 9$^{\prime}$ in the
Spitzer image. The  
field of view of LBC-Blue allows to cover it entirely and to sample also small portions of 
the M31 outer ring.
Seeing conditions during the observations were modest
(FWHM$\simeq 2\arcsec$ at 1.4 air-masses).
Nevertheless, we obtained an outstanding image of M31, and   
fully succeeded both in tracing the fine structure of the
innermost regions of Andromeda and finding new candidate stellar clusters in the galaxy.

The image was
bias-subtracted and flatfield-corrected using standard IRAF\footnote{IRAF is distributed by the
National Optical Astronomy Observatory, which is operated by the
Association of Universities for Research in Astronomy, Inc., under
cooperative agreement with the National Science Foundation.}/mscred procedures and best flat-field 
and bias images kindly made available by the LBC team\footnote{http://lbc.oa-roma.inaf.it/}.
Each chip was
then searched for sources down to $3 \sigma$ above the local background level 
using Source Extractor (Sextractor, Bertin \& Arnouts 1996). 
Sextractor provides
positions and magnitudes as well as a number of 
morphological parameters of the detected sources that can be used for source classification, 
e.g. to discriminate between Point Sources (PS) 
and Extended Sources (ES, see below). Our final list of well behaved sources
contains 5539 entries selected to have Sextractor quality flag equal to 0,
that means optimally measured sources.  


\begin{figure*}[htbp]
\centering
\includegraphics[width=15.4cm]{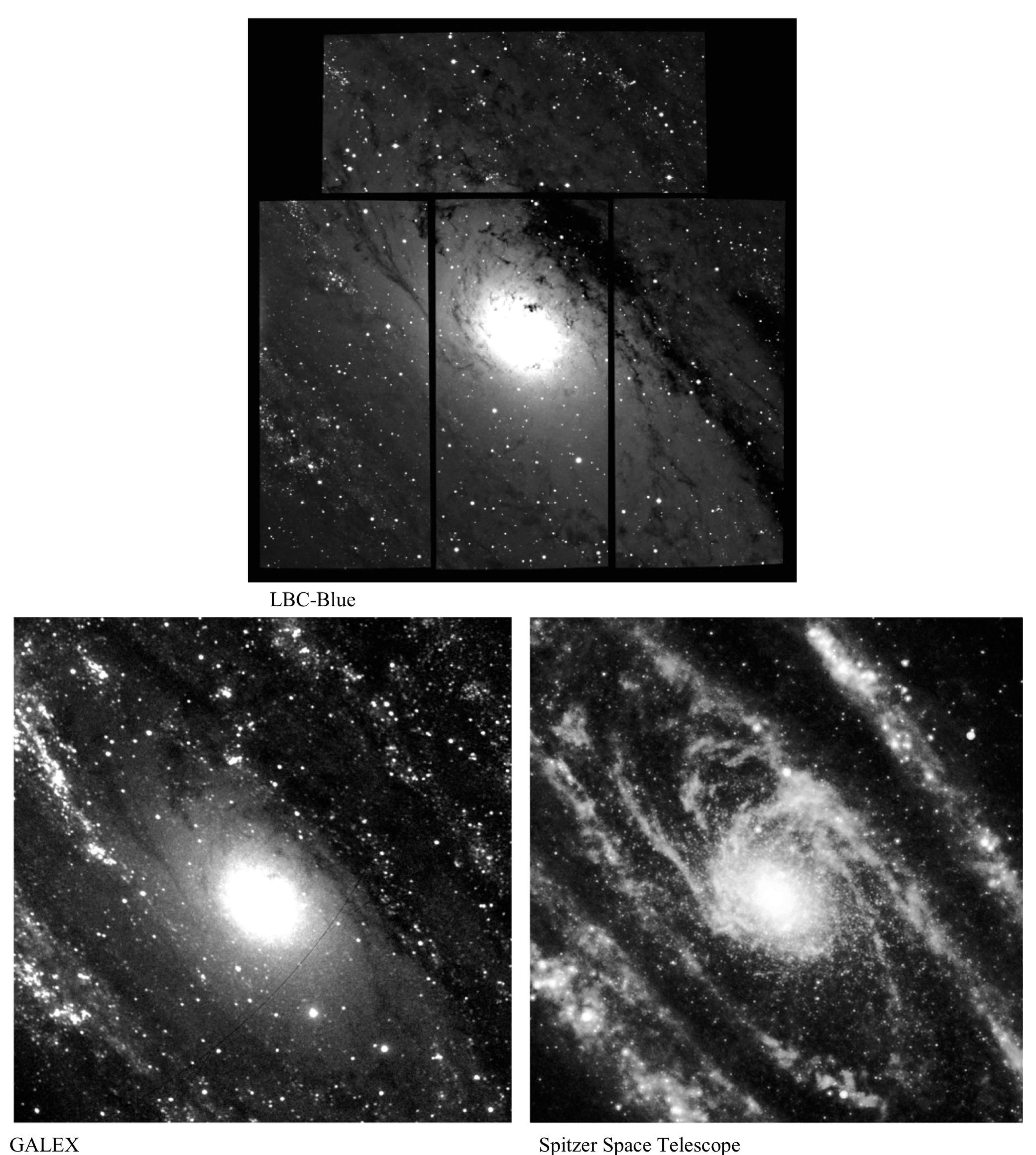}
\caption[angle=90]{\label{global}{\it Upper Panel}    
Sky-subtracted mosaic image of the center of M31 obtained with the LBC-Blue. 
{\it Lower-Left panel} The Andromeda galaxy observed with GALEX (from Thilker et al. 2005). The image is a composite
representation of the GALEX FUV and NUV data.  
Field of View is $\sim 24\arcmin \times 25\arcmin$, North is up and East is left. 
The M31 outer ring at a radius of 10 kpc is clearly visible in both images, instead the
inner ring is not resolved by the GALEX data, which only show diffuse light at its position.
{\it Lower-Right panel} The same portion of Andromeda observed with 
the  Multiband Imaging Photometer (MIPS) on board to the
Spitzer Space Telescope at a wavelength of 24 microns (from Gordon et al. 2006). 
Two well defined dust rings are clearly visible,
an outer ring at a radius of 10 kpc, and a second 1.5 by 1 kpc inner ring
offset by approximately 0.5 kpc from the galaxy nucleus.
Dimensions and orientation of the GALEX and Spitzer images are the same as 
in the LBC image.}

\end{figure*}

Astrometric catalogues typically used to search for counterparts 
and find astrometric solutions such as the GSC2 or the 2MASS have very 
few stars in this 
innermost region of M31. Fortunately,
a very appropriate secondary source is
the Johnson-Kron-Cousins UBVRI photometry down to $V\sim U\sim 23$ mag of 371781 sources in a $2.2$ deg$^2$ area of
M31 by Massey et al. (2006, hereafter M06). Our field is fully included within
the area covered by M06. M06 data-set contains stars as bright as $V\sim
14$ mag, that allow to obtain absolute astrometric solutions using
standards from the USNO-B1.0 catalogue (Monet et al. 2003).
We found  more than a thousand stars in common with M06 
in each chip of our image.
With these stars we derived astrometric solutions modelled by  third
order polynomials with residuals $\le 0.16\arcsec$ r.m.s. in both RA and
Dec. 
The astrometric solution was then ingested into the mosaic image
and a background subtracted version was produced using SWARP\footnote{See {\tt www.terapix.fr}}
(Bertin et al. 2002). This image is shown in the upper panel of 
Fig.~\ref{global}. We advise the reader that the electronic version of the
figure should be retrieved to appreciate the incredible amount of
information that is present in this image.
 
The stars in common with M06 also allowed us 
to derive a rough absolute photometric
calibration of the sources in our image. Since we lack 
colours we cannot apply colour-terms to our
calibration, thus the uncertainty in our  
absolute photometry is of the order of $\pm 0.5$ mag, very large but still
sufficient to provide a rough 
estimate of the limiting magnitude we have reached.
From 596 relatively bright ($U\la 20$ mag) stars in common between the two
catalogues we found $U=MAG\_AUTO+25.63$ mag, where MAG\_AUTO are the instrumental
magnitudes we obtained from Sextractor and U are the calibrated magnitudes in the
standard system from M06.
The faintest sources in our catalogue have $U\simeq 25$ mag, 90\% of the sources 
have $U\la 22$ mag . The uncertainty in the relative photometry for $U\la 22.0$ mag 
is $\epsilon_U\la 0.05$ mag, for $U\la 24.0$ mag is $\epsilon_U\la 0.10$ mag. 
This is perhaps the deepest wide-field, high resolution $U$ image of Andromeda ever obtained.

\section{Mapping the M31 dust lanes: LBC {\it vs} GALEX and Spitzer}

The LBT image (see upper panel in Fig.~\ref{global}) shows 
in great detail the fine structure of the dust lanes associated with the M31
inner spiral arms and provides a 
superb demonstration of the outstanding capabilities of the LBC-Blue system,
in a regime of high and strongly varying background such as the 
considered portion of Andromeda, where the very luminous and 
extended bulge is superposed to the inner
pattern of the spiral arms, traced by bright stars and prominent dust lanes.
In Fig.~\ref{global} we show the  LBC-Blue mosaic compared to a mosaic $\sim$ 24$^{\prime} \times 25^{\prime}$ 
image of the M31 nucleus obtained by GALEX (lower-left panel;Thilker et al. 2005) and
with an Infrared 
(24 $\mu m$) image (Gordon et al. 2006) of the Andromeda galaxy observed with the Multiband Imaging Photometer 
(MIPS; lower-right panel) on board 
to the Spitzer telescope\footnote{The image was retrieved from the NASA web site at 
{\tt http://ipac.jpl.nasa.gov/media\_images/ssc2005-20a1.jpg}}. 

The M31 outer ring at a radius of 10 kpc is clearly visible in the three images, while the
inner ring is not resolved by the GALEX data, which only show diffuse light at its position.
Indeed, the simple visual comparison of the LBC-Blue and the GALEX images allows one to appreciate 
the striking resolution capabilities of the blue camera on LBT. The structures of the M31 arms
that appear smeared and undefined in the GALEX image, are resolved into individual stars
by the LBC-Blue camera. The LBC-Blue image sharply outlines the contours of the M31 inner emitting regions 
that appear sparse and diffuse on the GALEX image.

The LBC-Blue and the Spitzer images are almost the positive and negative of each other, since  
at the infrared wavelengths sampled by Spitzer, dust is the main light emitter,
while in our U image we can see dust structures only because they intercept
the light coming from the disc and bulge of M31 lying in the background.
Hence, any dust lane or feature that is apparent in Spitzer images and is not
seen in the LBC $U$ image is probably located {\em behind} most of the stellar
disc of M31 in that direction. This comparison allows some insight on
the three-dimensional structure of the dust patterns in M31. 
On the other hand, a dust lane that is discerned on the LBC image but is not
seen in the Mid IR image may indicate that the feature has a different
temperature than the other structures. Moreover the finest details of the dust lanes 
and annuli can be discerned on the LBT image at a higher level of resolution than 
in the Spitzer image (Gordon et al. 2006).

A deeper discussion of these aspects  is
beyond the scope of the present paper, and here we simply draw the attention on another
rather straightforward use of the LBT image: a search for new candidate stellar clusters in M31.

\section{New candidate stellar clusters}

Since the review by Harris \& Racine (1979), there has been a longstanding debate in the literature about the
so called ``missing candidate" globular clusters in the bulge of M31 (see e.g. Barmby et al. 2001, Puzia et al. 2005, 
and references therein, for recent discussions of this problem). We have used our LBT $U$ image to search for new
globular clusters the inner regions of M31, altough  
the $U$ filter is far from being the ideal passband to search for
classical old (and hence red) GCs. In fact, our U image is best suited for the detection of young/hot stellar populations.
Therefore it is likely that any new cluster identified in this image is a young system, 
either an open cluster or a blue luminous compact cluster (BLCC) as recently found by Fusi Pecci et al. (2005).

The long experience in the search for GCs in M31 taught us that the lists of GC candidates 
are mainly contaminated by 
foreground Galactic stars and distant galaxies lying in the background, 
while HII regions and other types of spurious sources play a secondary role, (see
G04 and G06a).
Lacking high spatial resolution allowing to resolve clusters into individual stars (see Barmby
et al. 2007, Galleti et al. 2006b, 2007), the usual way to
disentangle GCs from background galaxies is via low resolution spectroscopy
allowing the measure of the source's radial velocity: distant galaxies
have high positive velocities due to cosmological recession, while {\it bona-fide} M31 GCs
should have velocities comparable to the galaxy systemic velocity, ($V_r(M31)\simeq -300 \pm 450$ km/s, G06a). 
For those cases where the actual nature of a candidate cannot be discerned by the
radial velocity, a crucial test is to check whether the target appears on the image as ES
(hence a genuine cluster) or as a PS (i.e. a star, see G06a). In the 
Revised Bologna Catalogue of the M31 globular clusters (RBC, G04) we keep
track not only of confirmed {\it genuine} GCs and of candidates yet to be
confirmed, but also list {\it false} candidates that turned out to be 
galaxies, stars and HII regions.

The high surface brightness background and the strong
extinction of the prominent dust structures in our LBT image proved a natural and very efficient 
shield against contamination by
background galaxies. Among the 124 spurious sources flagged as confirmed galaxies in the 
RBC, only one
occurs in the region sampled by our image. 
Moreover, the $U$ passband also disfavours contaminating background galaxies, which are typically quite
red (G04,G06a). Hence any ``roundish" source 
in our image that appears 
to be significantly more extended than a star of similar luminosity
may be a good stellar cluster candidate.  

To search for new candidates we 
first cross-correlated our catalogue against the RBC (which now includes also the clusters and candidates recently found by Kim et al. 2007), using 
a tolerance
radius of $1\arcsec$. In this way we recovered 62 confirmed
clusters, 27 confirmed stars, and 62 candidate clusters.
 
92\% of the M31 confirmed GCs recovered in our LBT image have $U$ magnitude brighter than 20 mag, hence  
about 5 magnitudes brighter than the limiting magnitude of our photometry ($U \sim$ 25 mag).
%
Then, we plotted all sources in a diagnostic diagram where  
the log of the isophotal area is displayed against the log of the isophotal
flux (log A - log F diagram, hereafter AF diagram, see G06a). 
This diagram is produced out of quantities directly output 
by Sextractor and is very effective in discriminating between PSs and 
ESs (see G06a).

   \begin{figure}
   \centering
   \includegraphics[width=9cm]{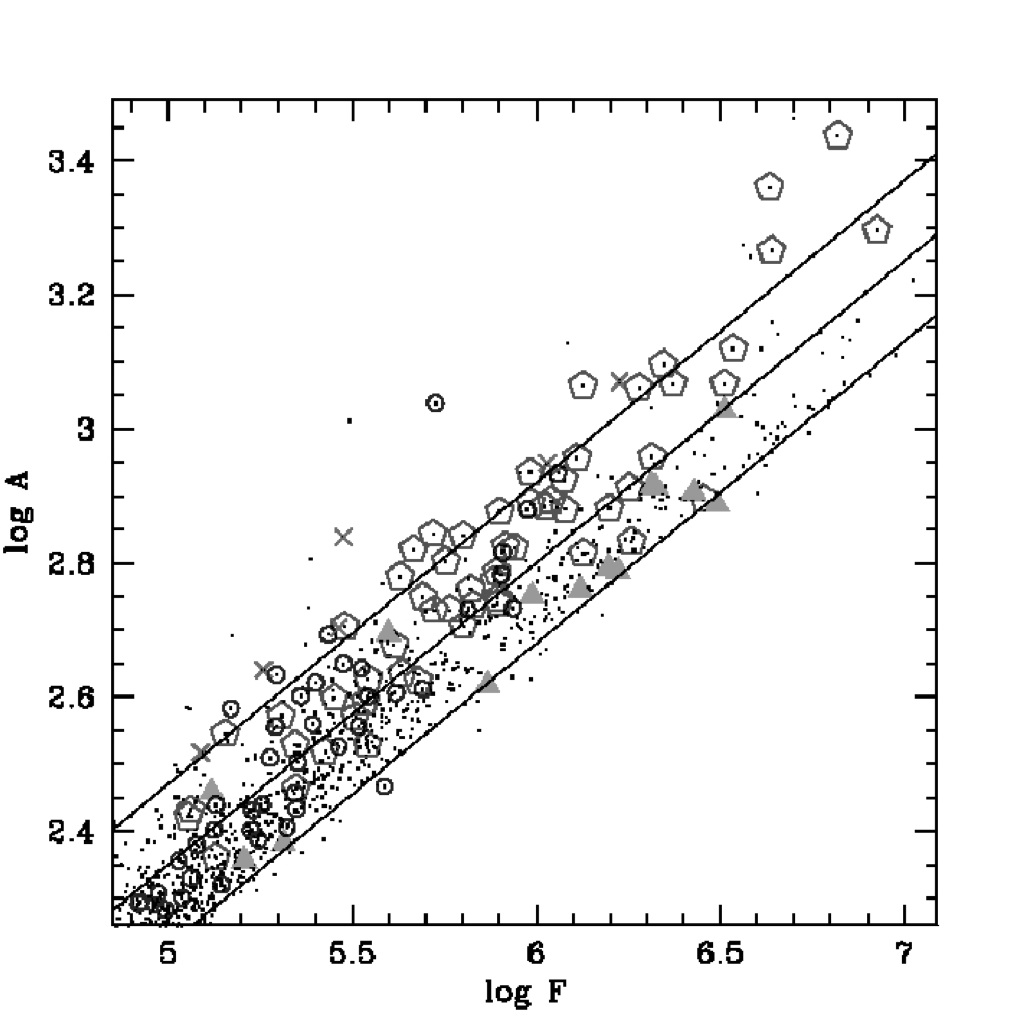}
    \caption{F-A diagram for the detected sources (dots). 
Pentagons are confirmed globular clusters, filled triangles are 
confirmed stars, and small open circles are the previously known candidate 
globular clusters from the RBC. Crosses are our new candidate 
clusters.
}
           \label{AF}
    \end{figure}
%
   \begin{figure}
   \centering
   \includegraphics[width=9cm]{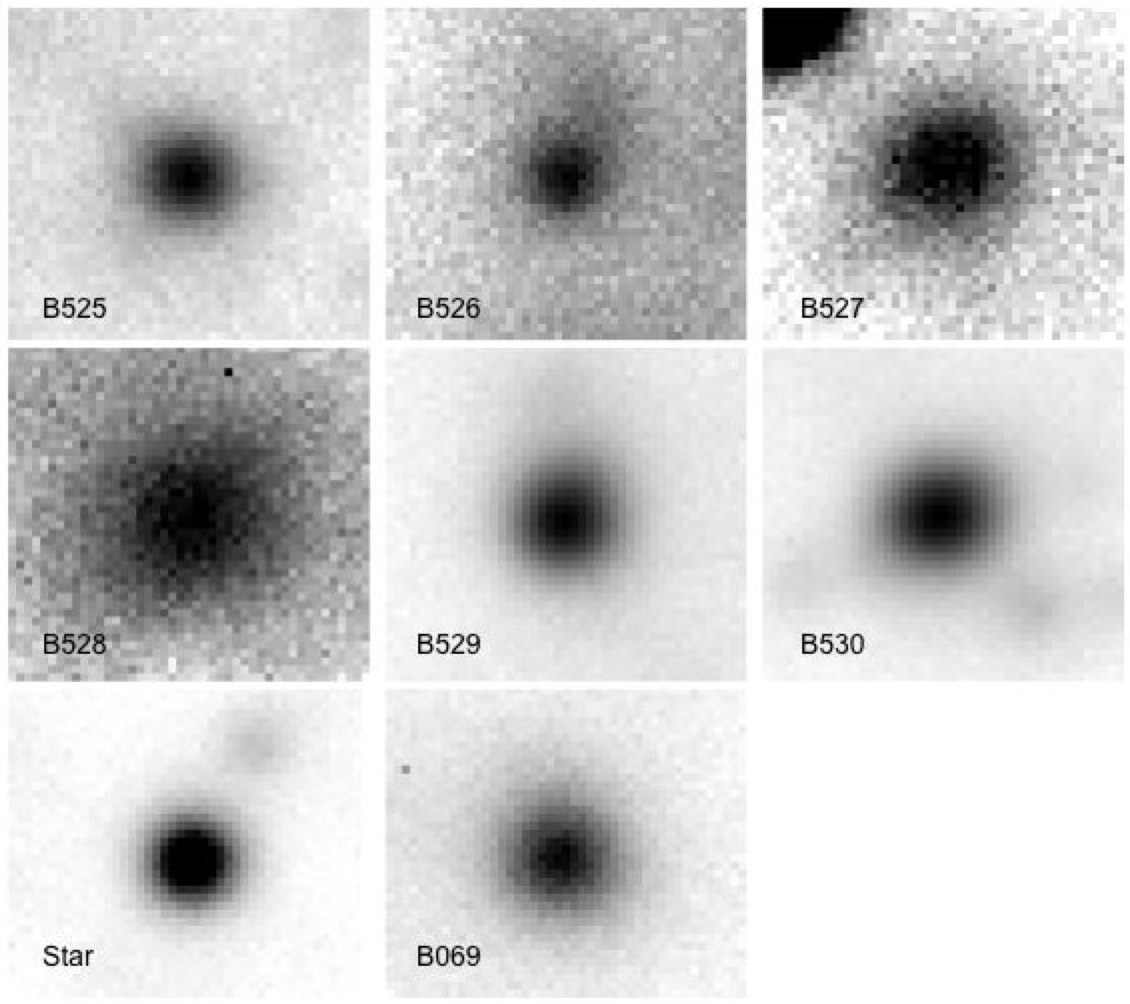}
    \caption{U images of the six new candidate clusters, a star 
(U$\sim$ 18.5 mag) and a confirmed globular cluster with U$\sim$ 18.8 mag (B069). Notice that B069
is one of the blue luminous compact clusters (BLCCs) recently found by Fusi Pecci et al. (2005).
Each image is $\sim10\arcsec$ on a side.  }
           \label{cand}
    \end{figure}
The AF diagram of the sources (small dots) in our LBT image is shown in Fig.~\ref{AF},
with confirmed GCs plotted as pentagons, previously known {\em candidate}
GCs as encircled dots, and the previously known (from the RBC) stars as filled 
triangles. 
The AF diagram is usually
dominated by a relatively narrow, well defined diagonal band with a sharp lower envelope
due to PSs (stars). Sources lying significantly above the PS
sequence are clearly extended (they have a larger isophotal area for their flux
with respect to a star). 
The poor seeing conditions of our observations resulted into a rather wide PS strip, as
judged by the position of known stars in the AF diagram. However, adopting 
a very conservative threshold and selecting as bona-fide ESs only sources lying above the upper
diagonal line running parallel to the PS band of Fig.~\ref{AF}, we were able to
identify 25 previously unknown Extended Sources.
Visual inspection of the image revealed that most of 
these new  ESs 
appear somehow
irregular or superposed to strong structures of the background and/or have 
ellipticity $e>0.4$ (see Barmby et al. 2007 and 
Galleti et al. 2007), or have some other problem making their
classification uncertain. Notice that the shape of each new ES was
locally compared to that of surrounding confirmed stars in order 
to exclude elongation effects due to worse focus
or astigmatism in some parts of the field.\\
To be conservative, we excluded all the uncertain sources, and were thus 
left with {\em six} well behaved, nearly round 
sources displaying a clear halo of light surrounding the central peak, as it 
is typical of genuine stellar clusters. 
Stamp size images of these six new good stellar cluster candidates 
are shown in Fig.~\ref{cand} along with one previously known genuine M31 GC
and one star of similar magnitude, for comparison.
The newly identified cluster candidates have 17.5 $\la U\la $19.5 mag; they have been named
according to the RBC convention, as they will be included in the updated
catalogue; their main characteristics  are summarised in 
Table~\ref{tab1}. Notice that, by comparing the U magnitudes of the new candidates with the values by M06,
we found  differences in the range from 0.3 mag for the brightest candidates up to 2.8 mag for the faintest ones,
with our measurements being systematically brighter. 
The reason is that we measured magnitudes of the extended sources by using isophotal fluxes, while M06 measured  
magnitudes by 
PSF fitting modelled on suitable PSF stars, thus probably losing light for the most extended sources.\\
To further investigate the nature of the new cluster candidates we checked the HST 
archive and found that B528 and B530 have counterparts on ACS@HST archival images.
B528 appears on three ACS images, namely: an F435W 2200 sec long exposure 
(Program 10006, PI Garcia), and on two 2370 sec long
exposures in the F606W and F814W filters
(Program 10260, PI Harris). On the F435W image the cluster appears as 
a young low-density association (probably an open cluster), that becomes progressively  
fainter on the F606W and F814W images, respectively. On the same frames we also 
identified two confirmed GCs, B162 and B159, which are 
bright and well defined on the ACS images while appear rather faint on the $U$ LBC-Blue image. B530 appears on 
an F606W and an F814W  WFPC2 3200 sec image
(Program 5971, PI Griffiths). In all the WFPC2 images the object appears as 
an extremely loose ensemble of stars.
 
We also computed the R parameter (i.e. the ratio of the FWHM of a given
source over the average FWHM of stars in the same image, G06a), where $R\gg 1.0$ values indicate
the extended nature of the source. Out of the 62 RBC candidate GCs found in our image, 19 are
found to have $R>1.20$, thus establishing their extended nature on a quantitative
basis. Their names and R ratios are provided in Table ~\ref{tab2}.
With this additional piece of information, any radial velocity estimate will suffice to confirm 
whether they are genuine M31 GCs or not.  

\begin{table}
\begin{minipage}[t]{\columnwidth}
\tiny
\caption{Newly discovered candidate Stellar Clusters}
\label{tab1}
\centering
\begin{tabular}{cccccc}
\hline \hline
 Name  & RA$_{J2000}$ & Dec$_{J2000}$ & U\footnote{Approximate values, see
 text} & FWHM  & R \\
\hline
B525 & 00:41:45.24 &41:12:29.38 &18.75 &  $2.5\arcsec$ & 1.22 \\ 
B526 & 00:42:03.08 &41:16:23.50 &18.90 &  $4.1\arcsec$ & 1.98  \\ 
B527 & 00:42:10.93 &41:29:59.02 &19.36 &  $4.3\arcsec$ & 2.05  \\ 
B528 & 00:43:11.90 &41:23:47.91 &19.25 &  $5.4\arcsec$ & 2.54  \\ 
B529 & 00:42:12.07 &41:25:38.66 &18.15 &  $2.6\arcsec$ & 1.22  \\ 
B530 & 00:42:50.31 &41:25:12.70 &17.72 &  $2.8\arcsec$ & 1.33 \\ 
\hline
\end{tabular}
\end{minipage}
\end{table}

\section[]{Discussion and Conclusions}

\begin{table}
\begin{minipage}[t]{\columnwidth}
\caption{R parameter for candidate M31 GCs from the RBC}
\label{tab2}
\centering
\begin{tabular}{cc|cc|cc}
\hline \hline
 Name  & R & Name & R & Name & R\\ \hline
  B079  & 2.113 &   B177  & 1.236 &B095D  & 1.206 \\  
  B080  & 1.590 &   B269  & 1.364 &B097D  & 1.480 \\  
  B108  & 1.277 &   B271  & 1.864 &AU008  & 2.382 \\
  B142  & 1.332 &   B057D & 1.353 & V234  & 1.449 \\
  B150  & 1.327 &   B072D & 1.279 & M001  & 1.225 \\    	 
  B157  & 1.231 &   B089D & 1.273 &	& \\	     	 
  B172  & 1.230	&   B092D & 1.873 &     & \\		  
\hline
\end{tabular}
\end{minipage}
\end{table}
The results presented in this paper provide first quantitative hints of the 
outstanding potential of the blue channel of the LBC on the Large Binocular Telescope.
%
In spite of rather poor seeing conditions, a very deep $U$ image of M31 was obtained, 
reaching $U\simeq 25.0$ mag and revealing a
harvest of details of the central region of Andromeda. Using a secondary
astrometric catalogue it was possible to derive a more than satisfactory global
astrometric solution (r.m.s.$\simeq 0.16^{\prime\prime}$). Though the background in the sampled region is very high and
strongly variable we were able to obtain relative
photometry accurate to $\pm$0.1 mag over the magnitude range $16.5\la U \la24.0$ mag.
Scientific return was easily obtained even from 
these very first commissioning images: 
(a) we highlighted fine structure details of the dust lanes in the 
inner regions of M31;
(b) six new good-quality M31 stellar cluster candidates were identified, to be followed up spectroscopically; 
(c) the extended nature of 19 previously known GC candidates was also established 
on firm quantitative basis.

%

\begin{acknowledgements}
Based on data acquired using the Large Binocular Telescope (LBT). The LBT is an
international collaboration among institutions in the United States, Italy and
Germany. LBT Corporation partners are: The University of Arizona on behalf of
the Arizona university system; Istituto Nazionale di Astrofisica, Italy; LBT
Beteiligungsgesellschaft, Germany, representing the Max-Planck Society, the
Astrophysical Institute Potsdam, and Heidelberg University; The Ohio State
University, and The Research Corporation, on behalf of The University of Notre
Dame, University of Minnesota and University of Virginia.
Cross-correlations between different data-sets and search for astrometric
solutions have been performed with the CataPack software, developed and
maintained by P.Montegriffo.
We thank the referee Pauline Barmby for suggestions and comments that have helped
to significantly improve the overall quality of the paper .
This research made use of the NASA/ADS database.
Financial support for this study was provided by INAF.

\end{acknowledgements}

\end{document}